\documentclass[aps,pra,twocolumn,superscriptaddress]{revtex4-2}
\usepackage{graphicx}
\usepackage{graphics}
\usepackage{dcolumn}
\usepackage{bm}
\usepackage{amssymb}
\usepackage{amsmath}
\usepackage{setspace}
\usepackage{hyperref}
\usepackage{epstopdf}
\usepackage{siunitx}
\usepackage{comment}
\usepackage{xfrac}
\usepackage{float}
\usepackage{upgreek} 
\usepackage[normalem]{ulem}
\usepackage[export]{adjustbox}
\hypersetup{%
   pdfpagemode=None, 
   pdfstartpage=1,
   pdfmenubar=true,
   pdftoolbar=true,
   colorlinks = true,
   linkcolor=blue,
   citecolor=blue,
   urlcolor=blue,
   bookmarksopen=false
 }

\begin{document}
\newcommand{\ket}[1]{|{#1}\rangle}

\title{A High-Fidelity Method for a Single-Step $N$-bit Toffoli Gate in Trapped Ions}



\author{Juan Diego Arias Espinoza}
\affiliation{Van der Waals-Zeeman Institute, Institute of Physics,
University of Amsterdam, 1098 XH Amsterdam, the Netherlands}
\author{Koen Groenland}
\affiliation{QuSoft Science Park 123, 1098 XG Amsterdam, the Netherlands}
\author{Matteo Mazzanti}
\affiliation{Van der Waals-Zeeman Institute, Institute of Physics,
University of Amsterdam, 1098 XH Amsterdam, the Netherlands}
\author{Kareljan Schoutens}
\affiliation{QuSoft Science Park 123, 1098 XG Amsterdam, the Netherlands}
\affiliation{Institute for Theoretical Physics, Institute of Physics, University of Amsterdam, Science Park 904, 1098 XH Amsterdam, the Netherlands}
\author{Rene Gerritsma}
\affiliation{Van der Waals-Zeeman Institute, Institute of Physics,
University of Amsterdam, 1098 XH Amsterdam, the Netherlands}

\date{\today}

\begin{abstract}

Conditional multi-qubit gates are a key component for elaborate quantum algorithms. In a recent work, Rasmussen {\it et al.} (Phys. Rev. A 101, 022308) proposed an efficient single-step method for a prototypical multi-qubit gate, a Toffoli gate, based on a combination of Ising interactions between control qubits and an appropriate driving field on a target qubit. Trapped ions are a natural platform to implement this method, since Ising interactions mediated by phonons have been demonstrated in increasingly large ion crystals. However, the simultaneous application of these interactions and the driving field required for the gate results in undesired entanglement between the qubits and the motion of the ions, reducing the gate fidelity. In this work, we propose a solution based on adiabatic switching of these phonon mediated Ising interactions. We study the effects of imperfect ground state cooling, and use spin-echo techniques to undo unwanted phase accumulation in the achievable fidelities. For gates coupling to all axial modes of a linear crystal, we calculate high fidelities ($>$ 99\%) $N$-qubit rotations with $N=$ 3-7 ions cooled to their ground state of motion and a gate time below 1~ms. The high fidelities obtained also for large crystals could make the gate competitive with gate-decomposed, multi-step variants of the $N$-qubit Toffoli gate, at the expense of requiring ground state cooling of the ion crystal.

\end{abstract}

\maketitle

\section{Introduction}

Quantum computers promise dramatic speedups in a variety of disciplines~\cite{Jordan2012,AspuruGuzik2005,Abrams1999,Jaksch2003,Lidar1999}, but remain challenging to scale up in practice. A major obstacle to executing elaborate quantum algorithms, is the need for gates that act conditionally on a large number of qubits. The prototypical example of such a gate is the $N$-qubit Toffoli gate, which flips a single `target' qubit if and only if all $N-1$ `control' qubits are in the state $\ket{1}$. Even though quantum devices with over 50 qubits have been reported~\cite{Zhang2017,Arute2019}, the largest Toffoli gate ever performed is, to our best knowledge, the case $N=4$~\cite{Figgatt2017}. This gap is surprising, because Toffoli gates (or equivalents) are essential ingredients of many basic computation steps, such as elementary arithmetic~\cite{Vedral1996,Cuccaro2004,vanMeter2005}, error correction~\cite{Paetznick2013}, and the Grover diffusion operator~\cite{Grover1996}.

Two different strategies exist to implement Toffoli gates. The first consists of decomposing a single $N$-qubit Toffoli gate into a circuit consisting of one- and two-qubit gates~\cite{Maslov2003,Shende2009,Linke2017} or multiqubit gates, such as the M\o{}lmer-S\o{}rensen gate in trapped ions~\cite{Sorensen1999,Maslov2018, Groenland2020}. The second approach is to perform the gate in a single step using interactions that are native to the specific platform~\cite{Isenhower2011,Khazali2020,Molmer2011,Rasmussen2020}. In particular, a recent proposal~\cite{Rasmussen2020} has demonstrated that by exploiting systems with an all-to-all Ising interaction in combination with a drive field on a single target qubit an $i$-Toffoli gate can be implemented. This gate differs only from the regular Toffoli by a phase $+i$ on the target qubit.

Trapped ions are a natural candidate to implement this proposal, as intrinsic Ising interactions have been demonstrated in increasingly large ion crystals~\cite{Sorensen1999,Leibfried2003,Roos2008,Kim2009,Zhang2017}. Moreover, quantum operations have been demonstrated~\cite{Ballance2016,Gaebler2016} with fidelities higher than 99.9\%. Ising interactions generally arise from qubit-phonon couplings $\hat{H}_\text{q-ph}$ generated from state-dependent laser-induced forces on the ions. Combining this mechanism with the driving field \(\hat{H}_\text{drive}\) required for an $i$-Toffoli gate poses a problem, as both process do not commute i.e. ~$[\hat{H}_\text{q-ph}, \hat{H}_{\text{drive}}]\neq 0$. As a result, the qubit states and the motion of the ions remain entangled at the end of the gate sequence, which leads to fidelity loss. This effect could be mitigated by restricting the strength of the spin-phonon coupling such that the phonons are only virtually excited~\cite{Kim2009}. However, limiting the strength of the Ising interactions leads to undesirably long gate times.

In this work, we show that this residual qubit-phonon entanglement can be suppressed by adiabatic ramping of $\hat{H}_\text{q-ph}$. In this way, the $i$-Toffoli gate operates on the dressed eigenstates of $\hat{H}_\text{q-ph}$, that are adiabatically connected to the Fock eigenstates of the non-interacting system. The benefit of this approach is that the effective Ising interaction strength does not have to be limited to the regime of virtual phonon excitation. We show that high-fidelity $\bar{F}>99\%$, single step, $i$-Toffoli gates should be possible with up to 7 ions at gate times $\sim$ 600~\(\upmu\)s.

We start in Sec.~\ref{sec:model} with the derivation of the model for a $N$-qubit $i$-Toffoli gate for a system of trapped ions and introduce our proposal for adiabatic preparation of dressed states. In Sec.~\ref{sec:linear_crystal} we analyze the results of numerical simulations for a linear 3 crystal and consider the role of inhomogeneous Ising interactions mediated by multiple phonon modes. We discuss the implementation of a method based on multi-frequency laser fields~\cite{Shapira2020} to eliminate undesired phases originating from these inhomogeneous  interactions. Finally, in Sec.~\ref{sec:fidelities} we calculate the fidelities for 3-9 qubits gates and discuss sources of errors and ways to mitigate them. We also consider the effects of imperfect ground state cooling.

\section{Model of a $N$-qubit Toffoli gate in trapped ions}\label{sec:model}

\subsection{Single step $N$-qubit $i$-Toffoli gate}

\begin{figure}
	\centering
	\includegraphics[width=0.5\textwidth,scale=1.0]{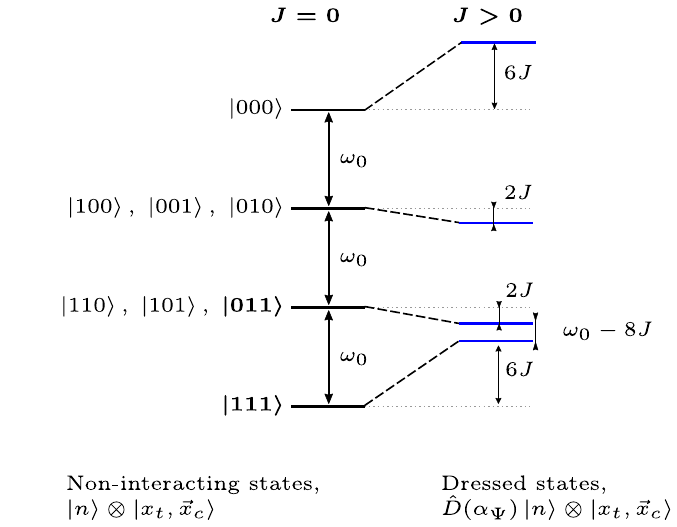}
	\vspace{10pt}
	\caption{\label{fig:Spectrum} Energies of non-interacting eigenstates (\(J=0\)) and interacting (dressed) states (\(J>0\)). The two target states \(\ket{111},\;\ket{011}\) are highlighted. Because their energy gap is unique, an appropriate drive field can couple the states resonantly.}
\end{figure}

Briefly, the proposal~\cite{Rasmussen2020} requires qubits coupled via an Ising interaction of the form $\hat{H}_\text{Ising} = \sum^N_{ij}J^{(i,j)}\hat{\sigma}_z^{(i)}\hat{\sigma}_z^{(j)}$ with \(\hat{\sigma}^{(i)}_r\) the Pauli matrix acting on ion $i$, and $J^{(i,j)}$ the strength of the interaction field \footnote{We define $\hbar=1$ and thus omit it from all the Hamiltonians in this text}. Including a drive field of frequency \(\omega_g\) with strength \(g\) acting on the target qubit, $\hat{H}_{\text{drive}}=g\hat{\sigma}_x^{(\text{t})}\cos{(\omega_g t)}$, and the energy of the non-interacting qubits, $\hat{H}_0=\omega_0/2\sum_i\hat{\sigma}_z^{(i)}$, a simple Hamiltonian is obtained:

\begin{equation}\label{eq:Htot}
    \hat{H}_\text{T} = - \frac{\nu}{2}\sum_i\hat{\sigma}_z^{(i)} + \sum_{i \neq j}J^{(i,j)}\hat{\sigma}_z^{(i)}\hat{\sigma}_z^{(j)} + \frac{g}{2}\hat{\sigma}_x^{(\text{t})},
\end{equation}
where we transformed into the interaction picture with respect to \(\omega_g\), using $\hat{U}=\exp\Big(-i \frac{\omega_g}{2}t\Big)$. We also define $\nu = \omega_g - \omega_0$ with $ \omega_0$ the energy spacing between qubit states (or eigenstates of $\hat{\sigma}_z$). These eigenstates and their energies (Fig.~\ref{fig:Spectrum}) can be labeled as $\ket{x_\text{t},\vec{x}_c}$ and \(E_{|x_t,\vec{x}_c\rangle}\) with $x_\text{t}$ describing the state of the target qubit and $\vec{x}_c$ is the string describing the state of the control qubits. In particular, the two target states labelled as $|0,1^{N_c}\rangle, |1,1^{N_c}\rangle$, where $N_c$ correspond to the number of control qubits, correspond to those that are coupled by the action of the Toffoli gate.

The driving field frequency (\(\omega_g\)) is chosen such that it resonantly couples these two states, i.e. $\Delta_{1^{N_c}} = E_{|0,1^{N_c}\rangle}-E_{|1,1^{N_c}\rangle} = \omega_g$. According to Eq.~\ref{eq:Htot} the energy gap for any pair of states with equal control bits can be written as:

\begin{align}\label{eq:E_gap}
	\Delta_{\vec{x}_c}=4\sum_{i=1}^{N_c} J^{(\text{t},i)}(-1)^{\vec{x}_i} + \omega_0,
\end{align}
where $x_i$ denotes the state of the $i$-th control qubit. The resonant condition becomes then \(\nu = 4\sum_{i=1}^{N_c} J^{(\text{t},i)}(-1)^{\vec{x}_i}\), which for the target states implies $\nu = -4\sum_{i=1}^{N_c} J^{(\text{t},i)}$.

Evolution under the Hamiltonian of Eq.~\ref{eq:Htot} for a (gate) time \(\tau_g=\pi/g\) leads to the desired $i$-Toffoli gate. To prevent accumulation of unwanted dynamical phases during the gate, timing restrictions can be considered, or an echo pulse can be applied. Both will be discussed later in this text.

\subsection{Implementation in trapped ions}

To achieve the required Ising interaction in trapped ions, a qubit state-dependent force is generated with two non-copropagating bichromatic lasers with beatnote frequency $\mu$, which excites phonons in the ion crystal. For an homogenous laser field extending over the full ion crystal, the laser-ion interaction Hamiltonian is \(\hat{H}_\text{q-ph} =  \sum_i F_i  \exp(i \vec{k}\cdot\hat{\vec{r}}^{\,(i)}) + \text{h.c.}\). Here \(F_i=(\Omega/2) e^{-i\mu t}\hat{\sigma}_z^{(i)}\) is a state-dependent interaction \footnote{The dependence on the qubit state in arises from a differential Stark shift set by proper choice of laser polarizations~\cite{Leibfried2003}} with $\Omega$ the interaction strength, \(\vec{k}\) the resulting wavevector of the interfering laser fields, and \(\hat{\vec{r}}^{\,(i)}\) the position operator of ion $i$. With \(\vec{k}\cdot\hat{\vec{r}}^{\,(i)}=\sum_m \eta_m^{(i)} ( \hat{a}^\dagger_m + \hat{a}_m ) \) the Hamiltonian can be written as:

\begin{equation}\label{eq:HI}
    \hat{H}_\text{q-ph} = \frac{\Omega}{2}\sum_{i}\left(e^{i\sum_{m}\eta^{(i)}_m\left(\hat{a}_m^{\dagger}+\hat{a}_m \right)-i\mu t} + \text{h.c.}\right)\hat{\sigma}_z^{(i)},
\end{equation}
where the creation and annihilation operators for the $m$-th phonon mode are denoted by $\hat{a}_m^\dagger$ and $\hat{a}_m$. The Lamb-Dicke parameter $\eta^{(i)}_{m}$ is scaled with the motion amplitude of the $i$-th ion on the $m$-th phonon mode ($\vec{b}^{\,(i)}_m$), i.e. $\eta^{(i)}_m= \vec{b}^{\,(i)}_m\cdot \vec{k} \sqrt{\hbar/(2M\omega_m)}$ with $M$ the ion mass and $\omega_m$ the phonon mode frequency.

Including again the drive field ($\hat{H}_\text{drive}$) and the energy of the non-interacting system (\(\hat{H}_0\)), the total Hamiltonian in the interaction picture of $\omega_g$ becomes:
\begin{align}\label{eq:Htot_ph}
\hat{H}_\text{T} =& -\frac{\nu}{2}\sum_i\hat{\sigma}_z^{(i)} + \sum_m \omega_m \hat{a}_m^\dagger\hat{a}_m \nonumber \\
+& \frac{\Omega}{2}\sum_{i}\left(e^{i\sum_{m}\eta_m^{(i)}\left(\hat{a}_m^{\dagger}+\hat{a}_m \right)- i\mu t} + \text{h.c.} \right)\hat{\sigma}_z^{(i)} \nonumber\\
+&  \frac{g}{2}\hat{\sigma}_x^{(\text{t})},
\end{align}
which includes a new (second) term for the motional energy of the system. Now the eigenstates of the non-interacting system have the form \({\ket{\Psi}}=\ket{\Phi}\otimes\ket{x_\text{t},\vec{x}_c}\), with \(\ket{\Phi}=\bigotimes_m \ket{n_m}\) the motional wavefunction of the system in the Fock space of the $m$ phonon modes of the crystal. For this system we define the target states for the $i$-Toffoli gate as the ones corresponding to an ion crystal cooled to its ground state, that is the two target states are \(\ket{\Psi_1}=\bigotimes_m \ket{n_m=0}\otimes\ket{1,\vec{x}_c}\) and \(\ket{\Psi_0}=\bigotimes_m \ket{n_m=0}\otimes\ket{0,\vec{x}_c}\) \footnote{In the following we will drop the motional component from states in its ground state and label them only by their electronic part, e.g. \(\ket{\Psi_0}\rightarrow\ket{0,\vec{x}_c}\)}.

Next, we simplify this Hamiltonian by going into the interaction picture of the phonon mode frequencies with the transformation $\hat{U}=\exp\Big(-i t\sum_m\omega_m \hat{a}_m^\dagger\hat{a}_m\Big) $: 
\begin{align}\label{eq:drive}
	\tilde{H}_\text{T} =&- \frac{\nu}{2}\sum_i\hat{\sigma}_z^{(i)} + \frac{\Omega}{2}\sum_{i} \Big{(}e^{i\sum_m \eta_m^{(i)}\left(\hat{a}_m^\dagger e^{i\omega_m t}+ \text{h.c.} -i \mu t\right)} \nonumber\\
	+& \text{h.c} \Big{)} \hat{\sigma}_z^{(i)} + \frac{ g}{2}\hat{\sigma}_x^{(\text{t})},
\end{align}
where high frequency terms ($2 \omega_g$) were ignored. We now consider a system within the Lamb-Dicke limit and transform the Hamiltonian into a new interaction picture \footnote{We use the rotating wave approximation and ignore frequencies higher than \(|\delta_\text{s}|\)} with respect to $\delta_\text{m} = \mu -\omega_m$ using $\hat{U}=\exp (-i t \sum_m \delta_m\hat{a}^\dagger_m \hat{a}_m)$:


\begin{align}\label{eq:HT_mm}
	\tilde{H}_\text{T,mm} =& -\frac{\nu}{2}\sum_i\hat{\sigma}_z^{(i)} +\frac{i\Omega}{2}\sum_m\sum_i\left(\hat{a}_m^\dagger -\hat{a}_m \right)\eta_m^{(i)}\hat{\sigma}_z^{(i)} \nonumber \\
	-&\sum_m\delta_m \hat{a}^\dagger_m \hat{a}_m +\frac{g}{2}\hat{\sigma}_x^{(\text{t})}.
\end{align}

To recover a Hamiltonian having the desired Ising interaction as in Eq.~\ref{eq:Htot}, we apply a Lang-Firsov transformation~\cite{Porras2004,Deng2005, Lang1968} to introduce a dressed-state picture of qubits entangled with phonon modes of the crystal. The transformation, \(\hat{U}_\text{I}=\exp \Big[-i\sum_{i,m} \alpha_m^{(i)}(\hat{a}_m^\dagger+\hat{a}_m)\Big]\), with \(\alpha_m^{(i)} = (\Omega\eta_m^{(i)}/2\delta_m)\hat{\sigma}_z^{(i)}\), has the form of a displacement operator that displaces the state of the system in phase space by a state dependent magnitude of \(\alpha_{m,\Psi} = \sum_i\alpha_m^{(i)}\). The result of the transformation is:

\begin{align}\label{eq:HT_ss}
	\tilde{H}_\text{T,I} = \hat{U}^\dagger_\text{I}\tilde{H}_\text{T,sm}\hat{U}_\text{I} = &-\frac{\nu}{2}\sum_i\hat{\sigma}_z^{(i)} +  \sum_{i\neq j} J^{(i,j)}\, \hat{\sigma}_z^{(i)}\hat{\sigma}_z^{(j)} \nonumber \\ &-\sum_m \delta_m \hat{a}^\dagger_m \hat{a}_m +\frac{\tilde{g}}{2}\tilde{\hat{\sigma}}_x^{(\text{t})},
\end{align}

\noindent with $J^{(i,j)}=\Omega^2 \sum_m\eta^{(i)}_m\eta^{(j)}_m/4\delta_m$, a corrected drive strength, \(\tilde{g}\), and a transformed drive term, \(\tilde{\hat{\sigma}}_x^{(\text{t})}=\hat{U}^\dagger_\text{I}\hat{\sigma}_x^{(\text{t})}\hat{U}_\text{I}\). Because the drive and the Ising terms do not commute, this transformation introduces a term \(\propto \alpha_m^{(\text{t})}\hat\sigma_y^{(\text{t})}\) which couples the drive to ion motion and can cause a gate error \(\propto \alpha_m^{(\text{t})}\). For weak (virtual) phonon excitation, \(\alpha_{\Psi}\ll 1\), such that \(\tilde{\hat{\sigma}}_x^{(\text{t})}\approx\hat{\sigma}_x^{(\text{t})}\), this error is small. However, this regime corresponds to very slow gates and we are here interested instead in the regime in which the corrections to \(\hat{\sigma}_x^{(\text{t})}\) have to be taken into account, i.e. \(\alpha_{\Psi} \gtrapprox  1\). 

The corrected drive strength \(\tilde{g}=g/\lambda^{\Psi^\prime,\Psi}_c\) accounts for the non-unitary overlap of the motional part of the (dressed) eigenstates of Eq.~\ref{eq:HT_ss}. These states are displaced Fock states, i.e. $|\Phi\rangle_\text{I}=\prod_m \hat{D}(\alpha_{m,\Psi})\ket{n_m}$, which can be produced adiabatically from the Fock states of the non-interacting system. The correction factor \(\lambda^{\Psi^\prime,\Psi}_c\) is equal to the overlap between the displaced states of any pair of states \(\ket{\Psi^\prime}, \ket{\Psi}\). The overlap is dependent on their initial phonon occupation number \(\ket{n_m}\) and can be written as~\cite{Cahill1969}:


\begin{align}\label{eq:lambda_c_mm}
	\lambda^{\Psi^\prime,\Psi}_c &= \prod_m \langle n_m^\prime|\hat{D}^\dagger(\alpha_{m,\Psi^\prime}) \hat{D}(\alpha_{m,\Psi})|n_m \rangle \nonumber \\ &= \prod_m e^{-\beta_m^2/2}\beta_m^{|\Delta n|_m} \left(\frac{n_m!}{n_m^\prime!}\right)^{\text{sign}(\Delta n_m)/2} L^{|\Delta n_m|}_{n_m}\left(\beta_m^2\right),
\end{align}
where \(\Delta n_m = n_m^\prime -n_m \) and \(\beta_m = \alpha_{m,\Psi^\prime} - \alpha_{m,\Psi}\), $L^{(\gamma)}_n(\beta)$ is the associated Laguerre polynomial. Note that the drive strength needed for implementing the correct gate depends therefore explicitely on the motional input state. For the target states in their ground states of motion, \(\ket{\Psi_0}, \ket{\Psi_1}\), the overlap simplifies to \(\lambda_c^{\Psi_0,\Psi_1} = \prod_m e^{-\beta_m^2/2}\) with \(\beta_m = \Omega \eta^\text{(t)}_m/\delta_m\) and where \(L_0\left(\beta_m^2\right)=1\).

\subsection{Adiabatic Preparation of States}

\begin{figure*}[ht!]%
	\includegraphics[width=0.9\textwidth,scale=1.0]{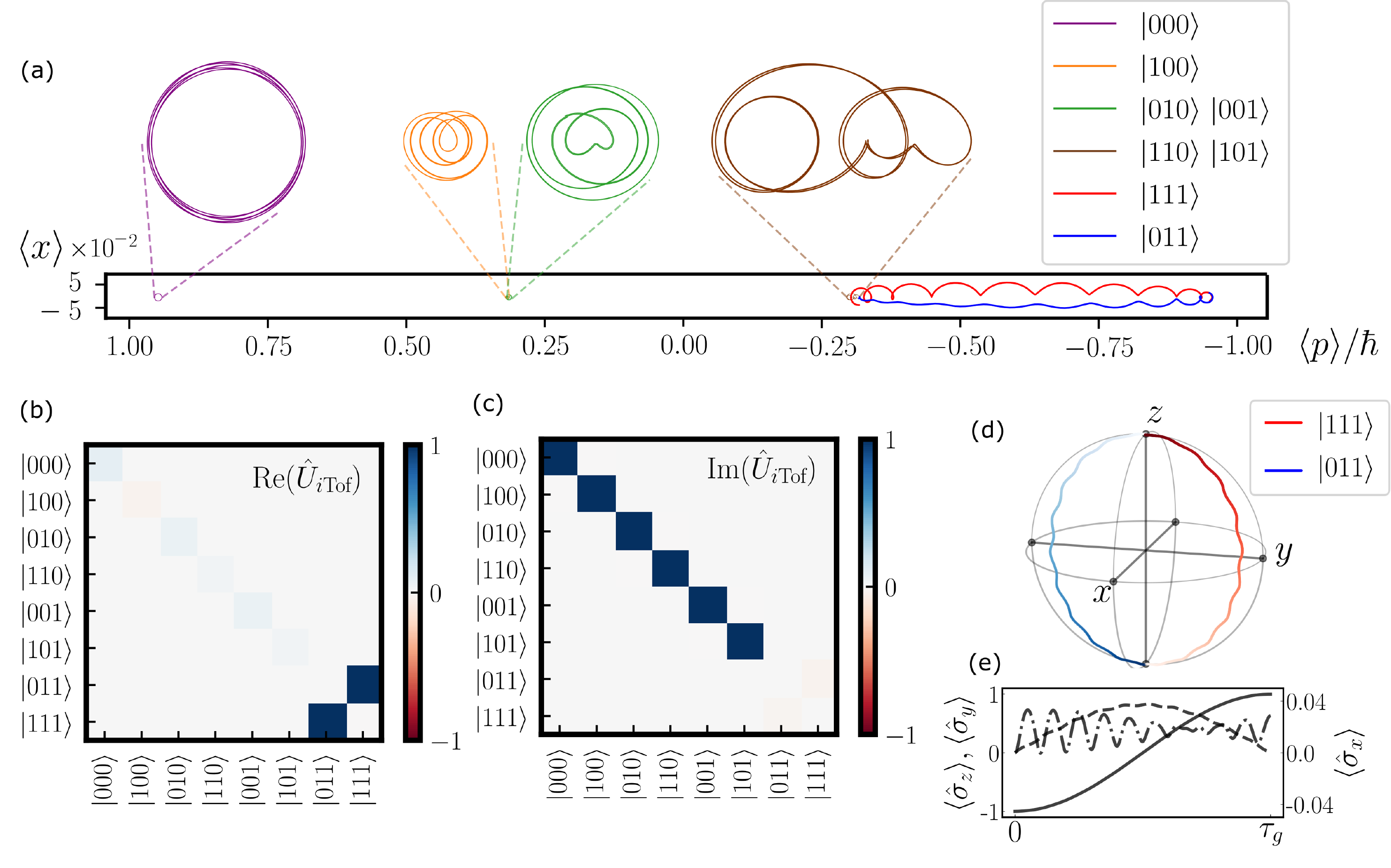}
  	\caption{\label{fig:3ionUnitaries} Time evolution of states under the action of \(\tilde{H}_\text{T,sm}\) (a) Phase space trajectories (zoomed in) of motional wavefunction during evolution with \(\hat U_\text{T}\). Note that the adiabatic ramp ensures that dynamics take place along the momentum axis in this frame, as explained more in detail in Appendix \ref{app:ASE}. (b) Real and (c) imaginary part of process unitary matrix for the motional ground state $\left(|n=0\rangle\right)$ subspace. (d) Evolution in the Bloch sphere of the two resonant states, and (e) the projections along x (\(\mathbf{-\cdot}\)), y (\(\mathbf{--}\)) and z (\(\mathbf{-}\)) of the trajectory of initial state \(|111\rangle\). Time is indicated with the color intensity from light ($t=0$) to dark ($t=\tau_g$) in (d). The gate parameters are $\delta_\text{CM}/2 \pi  = 20$ kHz, $J/2 \pi = 2$ kHz ($\Omega/2 \pi = 126.491$ kHz), $g/2 \pi = 1$ kHz for a gate time of \(\tau_g=\pi/g=500\;\upmu\)s.}
\end{figure*}

To guarantee a complete inversion of the target qubit, the system has to be prepared in a pure dressed eigenstate $|\Psi\rangle_\text{I}$ of the interacting system such that the drive strength can be exactly corrected using Eq.~\ref{eq:lambda_c_mm}. In the case of a sudden quench (diabatic activation) of Eq.~\ref{eq:HT_ss}, a superposition of dressed eigenstates will result. In contrast, by adiabatic switching (see Appendix \ref{app:ASE}) the qubit-phonon interaction, \(\hat{H}_\text{q-ph}\), and thus \(\hat{H}_\text{Ising}\), pure (dressed) eigenstates are obtained for which an appropriate drive strength can be chosen. 

It also makes our gate robust against residual phonon-qubit entanglement which in turn makes it less sensitive to timing errors. For quenched gates, this residual entanglement occurs if the total gate time \(t_\text{T} \neq 2k_1\pi/\delta_m \) (\(k_1 \in \mathbb{N} \)), as in this case the evolution of the states do not describe closed trajectories in phase space. In contrast, the adiabatic ramp assures that the system remains in an eigenstate during the laser-ion interaction. Therefore, the exact timing is not crucial as long as the ramp time is long enough to assure adiabaticity. In practice, however, setting \(t_\text{T} = 2k_1\pi/\delta_m\) still proves to be useful to reduce errors due to off-resonant drive field coupling between dressed states and to reduce errors caused by non-adiabaticity.

The gate sequence consists then in ramping up the interaction for a time \(t_a\) and performing the $i$-Toffoli gate (Eq.~\ref{eq:HT_ss}) for a time \(\tau_g\), and finally ramp down the interaction to transform the system back to the non-interacting or computational basis. This complete $i$-Toffoli process has a total length \(t_\text{T}=2 t_a + \tau_g\) and is described by:

\begin{equation}\label{eq:Tof_ad}
	\hat{U}_{i\text{Tof}} = \hat{U}^\text{d}_\text{eg}\hat{U}_\text{T}\hat{U}^\text{a}_\text{eg},
\end{equation}
where $\hat{U}^\text{a(d)}_\text{eg}$ is the unitary of the adiabatic activation (deactivation) of $\hat{H}_\text{Ising}$ and $\hat{U}_\text{T} = \exp(-i\tau_g \tilde{H}_\text{T,I})$.

\section{Simulations of a \emph{N}-qubit Toffoli gate in a linear ion crystal}\label{sec:linear_crystal}

\subsection{Single mode coupling}

The main features of our model can be first studied by considering an ideal system. This consists of a ground-state cooled linear ion crystal and an interaction laser coupling only to the axial modes of the crystal, with a beatnote $\mu$ tuned close to the center-of-mass phonon mode frequency $\omega_\text{CM}$ of the crystal, i,e. \(\delta_\text{CM}\ll\delta_{m \neq \text{CM}}\). We assume that the coupling with the remaining phonon modes can be ignored, i.e. \(J^{(i,j)}_\text{CM} \gg \sum_{m\neq \text{CM}}J_m^{(i,j)}\). This results in an homogeneous Ising coupling strength $J^{(i,j)} = \Omega^2 \eta_\text{CM}^2 / 4\delta_\text{CM} \equiv J $ and the simplified Hamiltonian:

\begin{align}\label{eq:CoM_Tof}
	\tilde{H}_\text{T,sm} = 2 N_c J\sum_i\hat{\sigma}_z^{(i)} + J \sum_{i\neq j} \, \hat{\sigma}_z^{(i)}\hat{\sigma}_z^{(j)} + \frac{\tilde{g}}{2}\tilde{\hat{\sigma}}_x^{(\text{t})} \nonumber \\ -\delta_\text{CM}\hat{a}_\text{CM}^\dagger\hat{a}_\text{CM}.
\end{align}

The resulting $i$-Toffoli process unitary for a 3-ion crystal is observed in Figs.~\ref{fig:3ionUnitaries}(b) and \ref{fig:3ionUnitaries}(c). We have chosen a ramp time (\(t_a\)) that ensures the adiabaticity of the process, and the disappearance of dynamical phases. These phases have the form \(\phi_{t_\text{T}} = \exp\left(-i E_{\ket{x_t,\vec{x}_c}} \tilde{t}_\text{T}\right)\), where the total effective process time is \(\tilde{t}_T = 2\tilde{t}_a + \tau_g\) and $\tilde{t}_a$ is effective ramp time (See Appendix~\ref{app:ASE}). Because the Ising couplings are homogeneous in this particular case, the phases vanish if $\tilde{t}_\text{T} J = 2k_2\pi$ (\(k_2 \in \mathbb{N}\)). For a modulation of the form \(\Omega(t<t_a)=\Omega \sin^2(\pi t/(2t_a))\) and these parameters both criteria are fulfilled by setting \(t_{a} = \tau_g \). 

To illustrate the dynamics under the action of Eq. \ref{eq:CoM_Tof}, we have plotted the phase space (Fig.~\ref{fig:3ionUnitaries}(a)) \footnote{The phase space shown in this work is in a rotating frame with frequency \(\mu\) and the values of \(\langle x\rangle\) are in units of the ground state wavepackage.} and Bloch sphere trajectories (Fig.~\ref{fig:3ionUnitaries}(d)) of the (target) dressed states $|\Psi\rangle_\text{I} = \hat{U}^\text{a}_\text{eg}\ket{n=0}\otimes |x_t,\vec{x}_c\rangle$. As expected for the two target states, the motional and electronic component are transformed from one to the other, i.e. \(\hat{D}(\alpha_{\Psi_0})\ket{n=0}\leftrightarrow\hat{D}(\alpha_{\Psi_1})\ket{n=0}\) and \(\ket{0,1^2}\leftrightarrow\ket{1,1^2}\). For the off-resonant states, closed trajectories are obtained indicating that motion is disentangled from the electronic component of the states. Finally, in Fig.~\ref{fig:3ionUnitaries}(e) we observe that the coupling of drive with the ion motion, leads to a small drive error reflected as small oscillations of \(\langle \hat\sigma_x\rangle\).

\subsection{Multi-mode coupling}

\begin{figure}[ht!]%
	\smallskip
	\includegraphics[width=0.48\textwidth,scale=1.0]{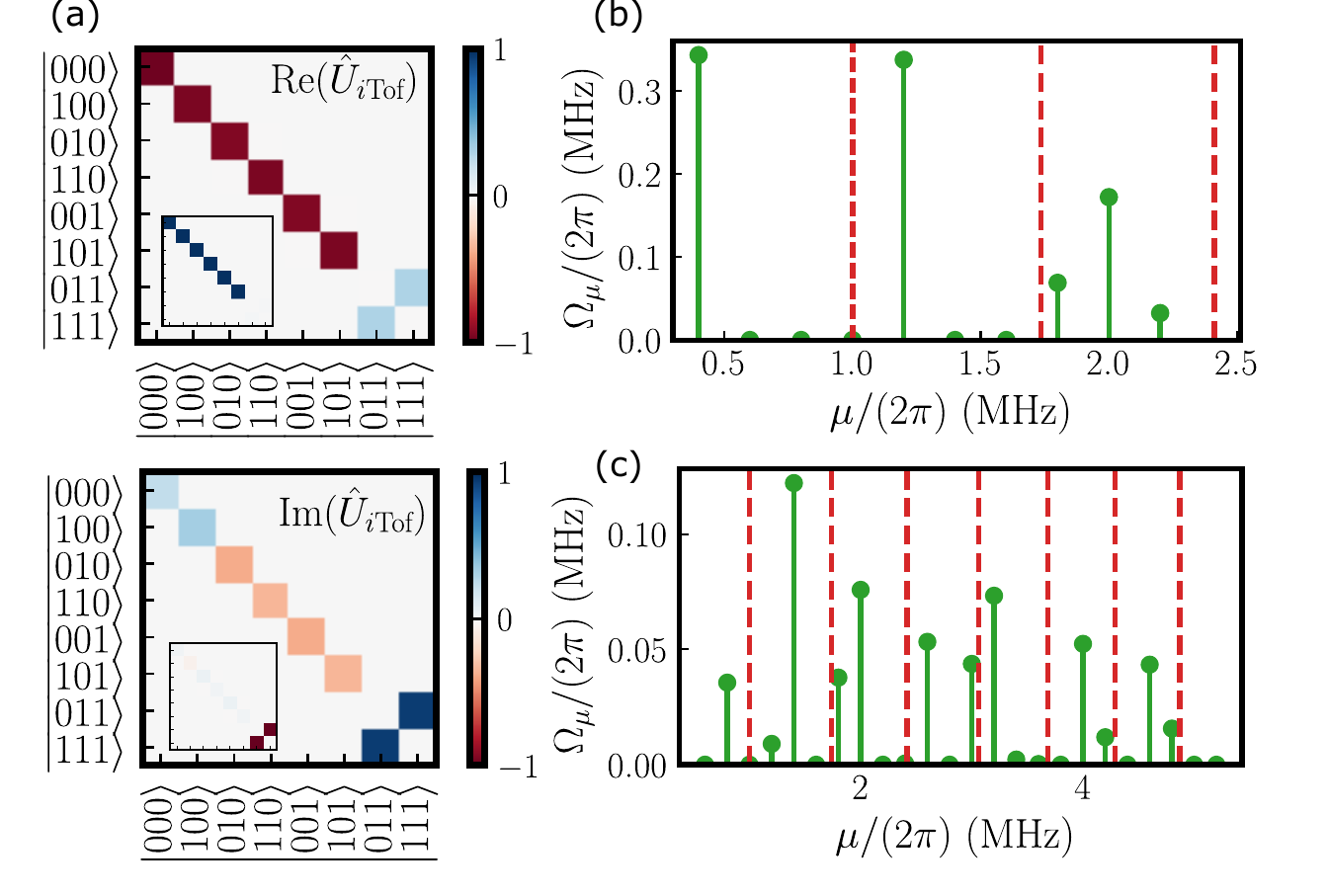}
  	\caption{\label{fig:MultiBeatnote} Multimode unitaries and spectrum of multiple beatnotes for ``echo'' step for phases cancellation. (a) $i$-Toffoli unitary for a 3 ion crystal considering all-mode couplings without and (inset) with ``echo'' step. Frequency and amplitude of beatnotes for (b) 3 and (c) 7 ions gate with detunings \(\delta_\text{CM}/2\pi=-20\) kHz and \(\delta_\text{CM}/2\pi=-50\) kHz respectively. The phonon mode frequencies are indicated in dashed red lines. The parameters of (a) are \(\omega_\text{CM}/2\pi = 1\) MHz, \(\delta_\text{CM}/2\pi=-20\) kHz, \(g/2\pi=1\) kHz) and for (b,c) the interaction time is \(t_\text{mb}\) 5 \(\upmu\)s.}
\end{figure}

In experiments, due to the finite spacing between phonon frequencies, the laser field will couple to multiple phonon modes, as described in Eq.~\ref{eq:HT_mm}. Although the dynamics of the gate will still be dominated by the coupling to the center-of-mass mode, the contributions of nearby modes, \(\sum_{m\neq \text{CM}}J_m^{(i,j)}\), will lead to two additional source of errors. The first are additional terms \(\propto \alpha_m^\text{(t)}\hat\sigma_y^\text{(t)}\) which increase the drive error, and the second are state-dependent dynamical phases. The latter occur because the Ising interactions are inhomogeneous, \(J^{(i,j)}\neq J^{(i,k)}\), thus the state energies are not longer proportional to a single value of $J$. As a consequence, no single gate time can be chosen such that they vanish at the end of the gate (Fig.~\ref{fig:MultiBeatnote}(a)). 

The first error can be minimized by using a linear crystals with odd number of ions and by addressing the central ion with the drive field. In this way, the largest contribution, coming from the next nearest phonon mode, disappears. To cancel the second error, dynamical phases are removed with an additional ``echo'' step. During this step, the sign of all coupling strengths is inverted \(J^{(i,j)}\rightarrow-J^{(i,j)}\) for a duration \(t_\text{T}\). To realize this echo, we follow a recent proposal~\cite{Shapira2020} in which a combination of multiple beatnotes coupling to all the phonon modes is used to generate couplings with arbitrary magnitude and sign. 

In short, the method uses beatnotes with frequencies \(\mu_k\) that are harmonics of the interaction time (\(t_\text{mb}\)) between the crystal and a multi-beatnote laser field, i.e. \(\mu_k= 2\pi k/t_\text{mb}\) for \(k \in \mathbb{N}\). Their amplitudes \(\Omega_{\mu_k}\) (Figs.~\ref{fig:MultiBeatnote}(b) and \ref{fig:MultiBeatnote}(c)) are calculated such that after a time \(t_\text{mb}\) the entanglement phases of each mode matches a target value \(\varphi_m\), and both dynamical phases and the entanglement with the phonon modes disappear. The entanglement phases are obtained by expressing the matrix of couplings for the echo step, \(\mathbf{\tilde{J}}_{i,j}=-J^{(i,j)}\), in terms of the phonon modes (\(\vec{b}_m\)) and the target entanglement phase:


\begin{equation}
	\mathbf{\tilde{J}} \approxeq \sum_{m=1}^N \varphi_m \vec{b}_m \otimes \vec{b}_m.
\end{equation}

To reduce the number of beatnotes required, we chose an interaction \(t_\text{mb} \sim 2k_1\pi/\omega_\text{CM}\) for a small integer \(k_1\), that also satisfies \(t_\text{T}=k_2 t_\text{mb}\) (\(k_2 \in \mathbb{N}\)). The ``echo'' is obtained by sequentially applying \(k_1\) multi-beatnote field pulses with the same modulation of the amplitudes \(\Omega_{\mu_k}\) as for the laser-ion coupling strength \(\Omega\) (See Appendix \ref{app:SE}). 




\section{Gate fidelities and error sources}\label{sec:fidelities}


\begin{figure}%
	\includegraphics[width=0.48\textwidth,scale=1.0]{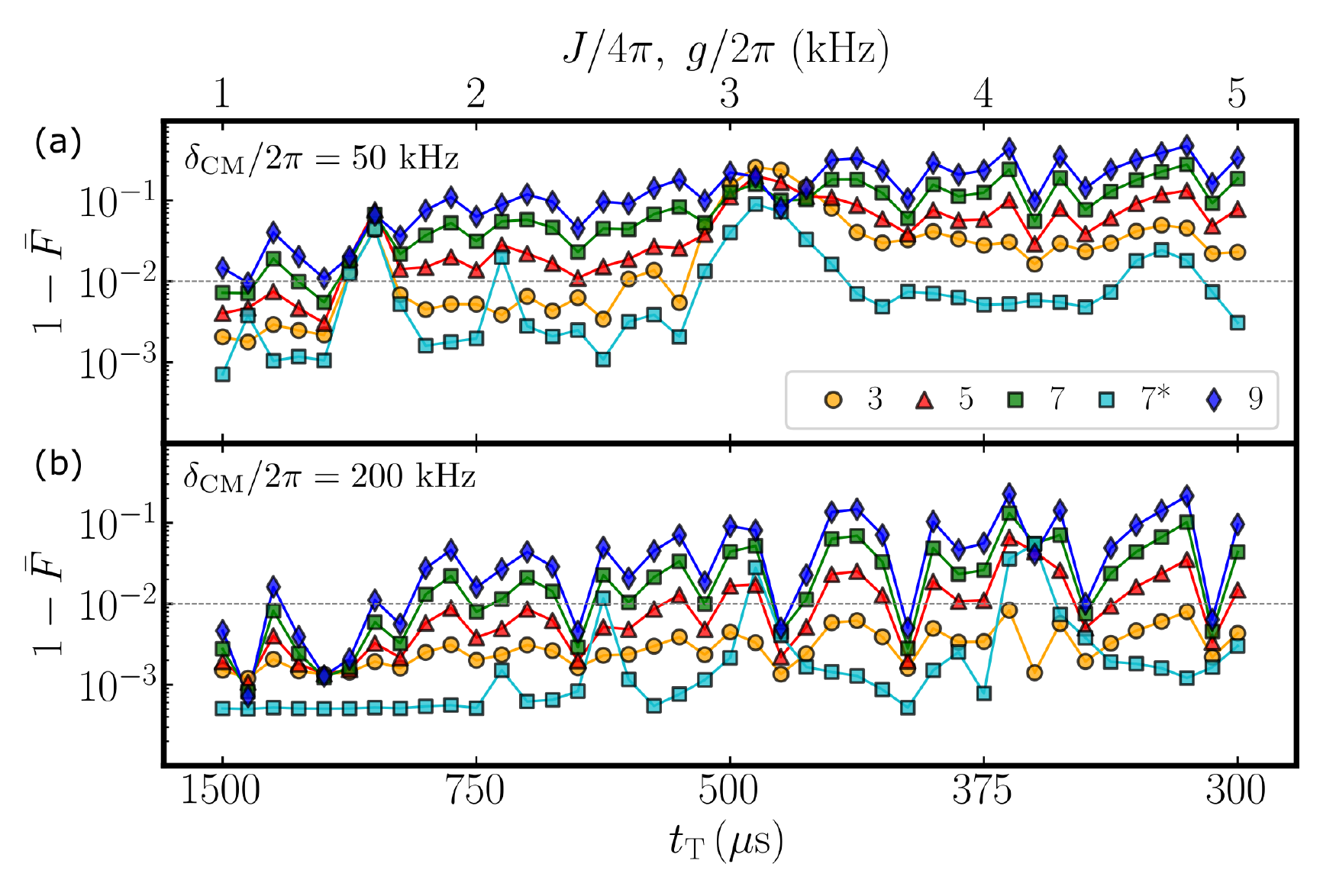}
  	\caption{\label{fig:FidvsSize_delta} Process error in function of Ising strength and gate time assuming single-mode coupling. The results are for a $i$-Toffoli gate of 3, 5, 7, and 9 ions for detunings of (a) 50,  and (b) 200 kHz. We also show the result (\(7^*\)) for a 7 ion crystal including an ``echo'' step. In this case, the total process time corresponds to \(2t_\text{T}\).}
\end{figure}

We have shown that an $i$-Toffoli gate ($\hat{U}_{i\text{Tof}}$) can be implemented in a linear crystal of ions in realistic conditions where the effective Ising interaction is generated by coupling to multiple phonon modes of the crystal. In this section, we will compare this gate against an ideal $i$-Toffoli gate ($\hat{U}_\text{Ideal}$) for different number of qubits and find conditions for fast gates with high fidelities. Additionally, we are interested in identifying and estimating the effect of other sources.

To characterize the gate, we use as figure-of-merit the average fidelity \(\bar{F}\)~\cite{Nielsen2002}:

\begin{equation}
	\bar{F}(\hat{U}_{i\text{Tof}},\hat{U}_{\text{Ideal}})=\frac{\sum_j \text{tr}[\hat{U}_{\text{Ideal}}U_j^\dagger\hat{U}_{\text{Ideal}}^\dagger\hat{U}_{i\text{Tof}}(U_j)]+d^2}{d^2(d+1)}
\end{equation}

\noindent where \(\hat{U}_{i\text{Tof}}(U_j)\equiv\text{tr}_\text{FS}\big(\hat{U}_{i\text{Tof}}[\hat{P}_0 \otimes U_j]\tilde{U}^\dagger_{i\text{Tof}}\big)\), \(U_j\) are generalized Pauli matrices in the qubit Hilbert space with dimension \(d=2^N\), \(\hat{P}_0=\bigotimes_m |0\rangle\langle 0|_m\) is a projector onto the $n_m=0$ Fock subspace and \(\text{tr}_\text{FS}\) is the partial trace of the phonons Fock space.

We start again by assuming single-mode coupling and calculate faster gates by increasing both $\Omega$ and $g$ and setting \(t_a=\tau_g\) to avoid phases accumulation. By increasing the interaction strengths and reducing gate and ramp times three types of gate errors will have to be accounted for: couplings between off-resonant states, drive errors and non-adiabatic couplings during ramping of the Ising interaction. To mitigate the first one, we require \(J > g\), therefore we keep the ratio $J/g=2$ for all the gates we will study. The last two errors can be minimized either by extending the duration of the adiabatic ramp or increasing the detuning of the laser beatnote \(\delta_m\), both reducing the amplitudes \(\alpha_{m,\Psi}\) and thus the final error. Because our goal is a faster gate, we have chosen for the latter.

Fidelities higher than 99\% with gate times below 500 \(\upmu\)s are obtained when \(\delta_\text{CM}/2\pi=200\) kHz (Fig.~\ref{fig:FidvsSize_delta}(c)) for gates with 3-9 qubits. As a consequence of the reduction of the ramp time with increasing $J$, the activation of the interaction becomes less adiabatic and the crystal motion is excited. This leads to coupling of motional excited states in the form of \(\ket{n>0}\ket{1,1^{N_c}}\leftrightarrow\ket{n>0}\ket{0,1^{N_c}}\) during the drive step. The larger drops in the fidelity are observed for particular interaction strengths, e.g. \(J/4\pi = 3.1\) kHz for \(\delta_t/2\pi=50\) kHz, originate also from undesired couplings between states of the type $|n=0\rangle\otimes|1,\vec{x}_c\rangle, \, |n=k\rangle\otimes|0,\vec{x}_c\rangle$, which become degenerate when \(\Delta_{\vec{x}_c}\sim k\delta_\text{CM}\). 

These errors affect more strongly gates with larger amount of qubits as the number of states and the occurrence of degeneracies increases. Furthermore, the drive and non-adiabaticity errors also increase, as the displacement amplitude \(\alpha_{m,\Psi} \propto N\). However, by choosing appropriate gate parameters, these undesired couplings can be avoided.


\subsection{Multi-mode coupling with residual crystal motion}

From the single-mode coupling analysis we have identified conditions for high fidelity gates for ion crystals in their ground state. We can use this information to calculate high-fidelity gates for systems where all axial phonon modes participate. We will also take into account residual ion motion such that average number of phonons in the crystal \(\bar{n}_m\) is not zero. In particular, we consider the cases where \(\bar{n}_\text{CM}>0\) and \(\bar{n}_\text{m $\neq $ CM}=0\). 

\begin{figure}[h!]%
	\centering
	\includegraphics[width=0.48\textwidth,scale=1.0]{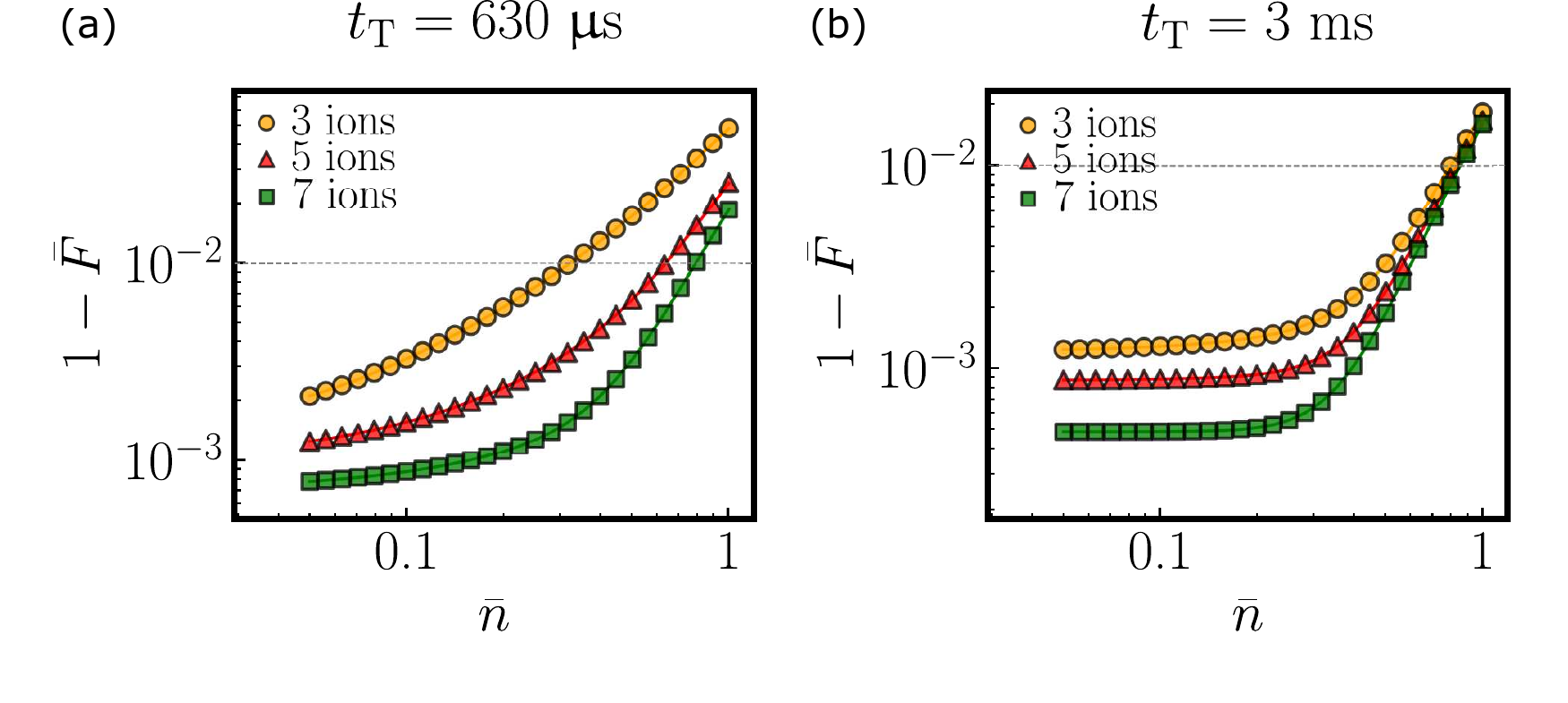}
 	\caption{\label{fig:FidvsN_ave} Effect of average phonon number in process fidelity for gate with multi-mode coupling. The Ising and drive strengths (\(J/4\pi=g/2\pi\)) are (a) 4.762 kHz and (b) 1 kHz. The detuning, \(\delta_\text{CM}/2\pi\), and the center-of-mass frequency, \(\omega_\text{CM}/2\pi\), are -200 kHz and 1 MHz respectively.}
\end{figure}

To illustrate, we choose gates with the largest detuning (\(\delta_\text{CM}/2\pi=-200\) kHz) to minimize drive errors and select two drive strength values (\(g/2\pi = 1.0;\; 4.762\) KHz) for which no large drop of fidelities were obtained in the single-mode model. As a result, we obtain multi-mode coupled gates with fidelities better than 99\% for both fast (Fig.~\ref{fig:FidvsN_ave}(a)) and slow gates (Fig.~\ref{fig:FidvsN_ave}(b)). Even in the presence of residual motion up to \(\bar{n}=1\), the fidelities always exceed 90\%.

Importantly, the addition of the ``echo'' step leads to fidelities that, in most of the cases, are better than those for single-mode model. Clearly, this step also compensates phases due to Stark shifts originated by couplings of states \(|1,\vec{x}_c\rangle \leftrightarrow |0,\vec{x}_c\rangle\), which remained uncorrected in Fig.~\ref{fig:FidvsSize_delta}.

Moreover, in absence of these phases, higher fidelities are obtained for larger gates (compare with Fig.~\ref{fig:FidvsSize_delta}). The increasing gaps between states, \(\Delta_{\vec{x}_c}\), for larger systems will reduce any type off-resonant couplings. In particular, it reduces couplings with excited motional states \(\Delta_{\vec{x}_c}\sim k\delta_\text{CM}\), as the ratio \(\Delta_{1^{N_c}}/\delta_m\) increases. Furthermore, not only do these gaps increase, there are also vastly more states with large gaps than with small gaps as $N$ increases. Thus state-specific errors weigh less in the calculation of the average fidelity for larger qubit gates.

\section{Discussion and conclusions}

We have presented a high-fidelity method to implement a single-step $i$-Toffoli gate in trapped ions. Our method allows operating in a regime of strong Ising interactions between qubits, necessary for fast gate operations. Although the adiabatic ramping of these interactions extends the total length of the process, the long coherence times offered by trapped ions~\cite{Wang2017} should allow the experimental implementation of this gate with high fidelities. Furthermore, recent methods of shortcut to adiabaticity~\cite{An2016, Baksic2016, Yan2019} may be applied to speed up the adiabatic preparation of states.

We have shown that, when the Ising interactions are mediated by multiple phonon modes, the residual dynamical phases can be effectively removed by using an ``echo'' step exploiting a recent non-adiabatic method for multiple qubit entanglement~\cite{Shapira2020}. A natural next step would be to combine our model and this method to generate homogeneous Ising interactions which should allows us to avoid the ``echo'' step.

A feature of our method is that the appropriate drive strength $\tilde{g}$ depends on the initial phonon state. Pure phonon input states can be assured by ground state cooling the ion crystal. The necessity of ground state cooling sets the implementation apart from a decomposition in e.g. M\o{}lmer-S\o{}rensen gates~\cite{Maslov2018,Groenland2020} that are more robust with respect to the phonon states~\cite{Sorensen1999,Kirchmair2009}. On the other hand, reaching the ground state via sideband cooling is an established technique in trapped ions and is used extensively.

Taking these considerations into account, our single step implementation of the $i$-Toffoli gate offers a competitive advantage compared to the gate-based decomposition, in particular for large $N$ when accumulated gate errors start to dominate.

\acknowledgements
We thank Georg Jacob for providing code for the multiple beatnote calculations, Arghavan Safavi-Naini, Philippe Corboz and Thomas Feldker for fruitful discussions. This work was supported by the Netherlands Organization for Scientific Research (Grant No. 680.91.120, R.G. and M.M.) and by the QM\&QI grant of the University of Amsterdam (K.G.).

\appendix

\section{Modulation of Ising interaction}\label{app:ASE}

\begin{figure}[ht!]%
	\includegraphics[width=0.5\textwidth,scale=1.0]{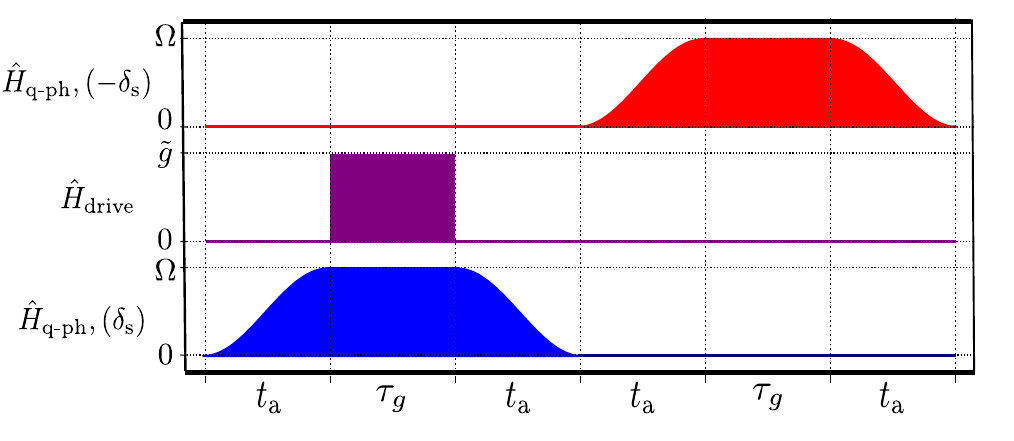}
	  \caption{\label{fig:Ramp_echo} Strength of Hamiltonian terms during length $i$-Toffoli gate. The Ising interaction (blue) is increased before acting with the drive field (purple) and then lower down again. A ``echo'' step (red) can be applied at the end of the gate to correct for residual entanglement or dynamical phases}
\end{figure}

The adiabatic transformation between the non-interacting and dressed states basis is realized by slowly increasing (decreasing) the strength of the Ising interaction for a time $t_\text{a}\gg 1/\delta_\text{s}$. This is achieved by modulating the Rabi frequency of the laser-ion Hamiltonian \(\hat{H}_\text{q-ph}\), such that \(\Omega(t)=\Omega \sin^2\big(\tfrac{\pi}{2} t/t_a\big)\) for \(t<t_a\) and \(\Omega(t^\prime)=\Omega \cos^2\big(\tfrac{\pi}{2} t^\prime/t_a\big)\) with \(t^\prime = t- t_a - \tau_g\) for \(t > t_a + \tau_g\) (Fig.~\ref{fig:Ramp_echo}). As a result we obtain the time-dependent Ising couplings $J(t)\propto \Omega(t)^2$. This modulation leads to a pulse area equivalent to that of a square pulse of half the length, such that we define an effective ramp time as \(\tilde{t}_a=0.5 t_a\).

As seen in Fig.~\ref{fig:PS_ramp}(a)-\ref{fig:PS_ramp}(b), the displacement in phase space of the two target states are significantly reduced for the adiabatically initialized system. This minimizes errors due to the non-commutativity between the drive and Ising interaction fields and also the ones arising from residual phonon-qubit coupling.
To approximate the unitary evolution of this adiabatic process we use a Trotter-Suzuki expansion \footnote{$[\hat{H}_\text{Ising},\hat{H}_0] \neq 0$, $\hat{U}^\text{a(d)}_\text{eg}$}:

\begin{figure}%
	\includegraphics[width=0.48\textwidth,scale=1.0]{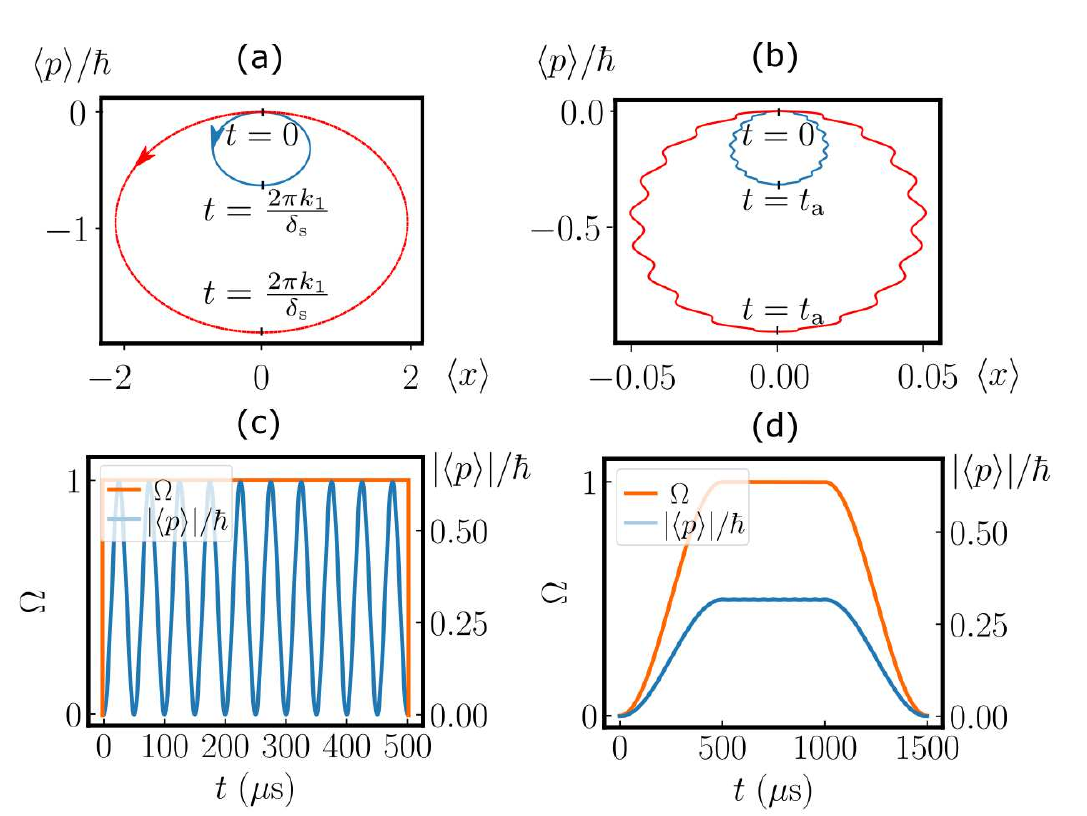}
  	\caption{\label{fig:PS_ramp} Evolution of target states under the application of \(\hat{H}_\text{q-ph}\). Trajectories of \(\ket{111}\) (red) and \(\ket{011}\) (blue) wavepackages and evolution of momentum expectation value of \(\ket{011}\) due to (a,c) a quench activation of 500 $\upmu$s and (b,d) an adiabatic modulation of \(\hat{H}_\text{q-ph}\)}
\end{figure}

\begin{align}
    \hat{U}^\text{a}_\text{eg} &= \prod^{t_\text{a}}_{t=0} e^{-i \Delta t \hat{H}_\text{Ising}(t)}e^{-i \Delta t \tilde{H}_0} \nonumber\\
    \hat{U}^\text{d}_\text{eg} &= \prod^{0}_{t=t_\text{a}} e^{-i \Delta t \hat{H}_\text{Ising}(t)}e^{-i \Delta t \tilde{H}_0}
\end{align}
where,

\begin{align}
	\hat{H}_\text{Ising}(t) &=  J(t)\sum_{i\neq j} \, \hat{\sigma}_z^{(i)}\hat{\sigma}_z^{(j)}\\
	\tilde{H}_0 &= 2 N_c J(t_a)\sum_i\hat{\sigma}_z^{(i)} -\delta_\text{s}\hat{a}_\text{s}^\dagger\hat{a}_\text{s}
\end{align}
and $\Delta t \ll (1/k)\delta_\text{s} \ll t_\text{a}$ is the time-step of the expansion and \(k=t_\text{a} \delta_t/2\pi\).

\section{Elimination of residual entanglement and dynamical phases}\label{app:SE}

Whenever the timing condition for the elimination of dynamical phases, \(\tilde{t}_\text{T} J = 2k_2\pi\), is not fulfilled, it is possible to add an additional ``echo'' step to the process to correct for these errors (Fig.~\ref{fig:Ramp_echo}). In this step the sign of the interaction strength is also reversed, i.e. \(J \rightarrow -J\). For the single mode coupling model, this is obtained by inverting the sign of the detuning\(\delta_\text{s} \rightarrow -\delta_\text{s}\). In the case of multi-mode coupling, we have used a combination of multiple beatnotes to generate an effective Ising interaction reversing the sign of the couplings during the gate step. More details of this method can be found in \cite{Shapira2020}.

The modulation of the coupling strengths between the ion and the single \(\Omega\) or multi-mode laser fields \(\Omega_{\mu_k}\) is equal to the one during the application of the initial gate. Furthermore, the length of the step needs also to be equal to total process time \(t_\text{T}\) and during the constant coupling strength portion of the ``echo'', no drive field is applied. In summary, this step can be described by the unitary:
\begin{equation}
	\hat{U}_\text{SE}=\hat{U}^\text{a}_\text{eg}\hat{U}_\text{I}\hat{U}^\text{d}_\text{eg},
\end{equation}
where \(\hat{U}_\text{I} = e^{-i\tau_g\big(\hat{H}_\text{Ising}(t_a)+\tilde{H}_0\big)}\) and the signs of $J(t)$ and $\delta_\text{s}$ are inverted in the Hamiltonians \(\hat{H}_\text{Ising}(t)\) and \(\tilde{H}_0\). 

\end{document}